\documentclass[twocolumn,aps,prc,floatfix]{revtex4}
\usepackage{graphicx}
\usepackage{dcolumn}
\usepackage{bm}
\usepackage[version=4]{mhchem}
\usepackage{color}
\def\es0{$E_{\rm sym}(\rho_0)$}

\def\us0{$U_{\rm sym}(\rho_0,k_F)$~}

\def\l0{$L(\rho_0)$~}

\begin{document}

\title{Imprints of High-Density Nuclear Symmetry Energy on Crustal Fraction of Neutron Star Moment of Inertia}

\author{Nai-Bo Zhang$^{1}$\footnote{naibozhang@seu.edu.cn} and Bao-An Li$^2$\footnote{Bao-An.Li@Tamuc.edu}}

\affiliation{$^1$School of Physics, Southeast University, Nanjing 211189, China}
\affiliation{$^2$Department of Physics and Astronomy, East Texas A$\&$M University, Commerce, TX 75429-3011, USA}
\date{\today}

\setcounter{MaxMatrixCols}{10}

\begin{abstract}
The density dependence of nuclear symmetry energy $E_{\rm sym}(\rho)$ remains the most uncertain aspect of the equation of state (EOS) of supradense neutron-rich nucleonic matter. Utilizing an isospin-dependent parameterization of the nuclear EOS, we investigate the implications of the observational crustal fraction of the neutron star (NS) moment of inertia $\Delta I/I$ for the $E_{\rm sym}(\rho)$. We find that symmetry energy parameters significantly influence the $\Delta I/I$, while the EOS of symmetric nuclear matter has a negligible effect. In particular, an increase in the slope $L$ and skewness $J_{\rm sym}$ of symmetry energy results in a larger $\Delta I/I$, whereas an increase in the curvature $K_{\rm sym}$ leads to a reduction in $\Delta I/I$. Moreover, the $\Delta I/I$ is shown to have the potential for setting a lower limit of symmetry energy at densities exceeding $3\rho_0$, particularly when $L$ is constrained to values less than $60$ MeV, thereby enhancing our understanding of supradense NS matter.
\end{abstract}
\maketitle


\section{Introduction}

The equation of state (EOS) of neutron-rich matter plays a crucial role in determining many properties of neutron stars (NSs). Unfortunately, it still has significant uncertainties, especially at high densities. The~high-density behavior of nuclear symmetry energy $E_{\rm sym}$ has long been recognized as the most uncertain aspect of the EOS of supradense neutron-rich nucleonic matter. This uncertainty stems from our limited understanding of the weak isospin dependence of the strong force, the~spin--isospin dependence of three-body nuclear forces, and~the tensor-force-induced isospin dependence of short-range nucleon--nucleon correlations in dense matter and is compounded by the challenges of accurately solving nuclear many-body problems \citep{Xu10}. The~symmetry energy significantly influences various properties of neutron stars, including their radii, deformations and polarizabilities \citep{Plamen}, and moments of inertia, and~the crust--core transition density \citep{Lattimer00, Lattimer01}. Thus, constraining the EOS or $E_{\rm sym}$ remains a critical issue for both the nuclear physics and astrophysics \mbox{communities~\citep{Danielewicz02,Ebook, Lattimer16, Watts16, Oertel17, Ozel16, Blaschke18}.}

Fortunately, terrestrial nuclear experiments and astrophysical observations have provided tighter constraints over the past decades. For~example, the~mass of the most massive NS observed so far, e.g.,~the mass of PSR J0740+6620, has recently been updated to \mbox{$2.08\pm0.07$ M$_\odot$ \citep{Cromartie19, Fonseca21}}, with~its radius constrained to $R=13.7^{+2.6}_{-1.5}$ km \citep{Miller21} or \mbox{$R=12.39^{+1.30}_{-0.98}$ km \citep{Riley21}} by the Neutron Star Interior Composition Explorer (NICER) Collaboration. Additionally, with~a significant increase in the dataset, updated analyses of the radius have been reported in refs. \citep{Salmi22, Salmi24, Dittmann24}. Furthermore, the~tidal deformability of a neutron star with 1.4 M$_\odot$ (canonical neutron star) has been extracted from the binary neutron star merger event GW170817, yielding $80<\Lambda_{1.4}<580$ at the 90\% confidence level, as~reported by the LIGO and Virgo Collaborations \citep{LIGO18}. Although~these observations still carry substantial uncertainties, they have been repeatedly employed in various analyses to enhance our understanding of neutron star matter (see, e.g.,~refs. \citep{Baiotti19, Burgio21, Li21} for reviews).

In addition to the aforementioned observational data, the~glitch phenomenon in pulsars can also serve to constrain the EOS or $E_{\rm sym}$ \citep{Hooker15, Newton15, Liu18, Dutra21, Parmar22}. This phenomenon, characterized by sudden increases in spin frequency, was first observed in the Vela pulsar \citep{Radhakrishnan69, Reichle69} and has since been detected in approximately 6\% of all known pulsars. Glitches occur due to the transfer of angular momentum from the superfluid component of the stellar interior to the solid crust \citep{Anderson75, Ruderman76, Pines85}. Previous studies have indicated that the glitch phenomenon is closely associated with the crustal fraction of the moment of inertia $\Delta I/I$, where $I$ represents the total moment of inertia of the star and $\Delta I$ is the moment of inertia of the crust. By~combining glitch data from the Vela pulsar and six other pulsars, ref. \citep{Link99} established that $\Delta I/I$ in a neutron star must exceed 1.4\%. Notably, this constraint does not depend on the mass of the neutron star, leading many references to assume masses of either 1.0 M$_\odot$ or 1.4 M$_\odot$, resulting in conditions of $(\Delta I/I)_{1.4}\geq1.0\%$ or $(\Delta I/I)_{1.4}\geq1.4\%$. This lower limit was later increased to 1.6\% \citep{Espinoza11}. While this limit can be met by nearly all EOSs, accounting for the entrainment of superfluid neutrons in the crust raises the constraint to $\Delta I/I\geq7\%$ \citep{Andersson12, Chamel12}, providing a more stringent condition for constraining the EOS or $E_{\rm sym}$.

Based on the constraints inferred from analyzing the glitch phenomenon, previous studies have found that a strong correlation exists between the density dependence of symmetry energy and the crustal moment of inertia (see, e.g.,~\citep{Dutra21,Fattoyev10,Liu18}), especially within the framework of microscopic or phenomenological nuclear many-body theories. However, no clear constraints on the EOS or $E_{\rm sym}$ have been extracted. On~the other hand, with~the accumulation of more of the NS observations mentioned above, many parameterized EOSs (see, e.g.,~refs. \citep{Read09,Margueron17a,Margueron17b,Annala18,Zhang18}) have been proposed and found very useful due to their flexibility and convenience in generating vastly different EOSs continuously. While not necessarily directly connected with a specific force within a specific many-body theory, such meta-model EOSs have been found very useful in extracting constraints on the EOS or $E_{\rm sym}$ within either Bayesian analysis \citep{Xie19} or forward modeling \citep{Zhang18}. Moreover, results from such an analytical formula and simple physics pictures are consistent with those starting from using different microscopic/phenomenological methods 
 with various interactions (see, e.g.,~ref. \citep{Zhang19}, for~detailed~comparisons).

In our previous study \citep{Zhang2021}, we simultaneously considered the maximum observed mass of PSR J0740+6620, the~simultaneous measurement of mass and radius for this pulsar, and~the tidal deformability from GW170817 to extract constraints on the EOS and $E_{\rm sym}$. We found that the $E_{\rm sym}$ at densities greater than $3\rho_0$ still exhibits significant uncertainties, highlighting the need for additional observational data. In~this work, we examine prospects of using the crustal fraction of NS moment of inertia $\Delta I/I$ to constrain the EOS and especially $E_{\rm sym}$ at suprasaturation densities using an isospin-dependent parameterization for the EOS of NS matter.

The remainder of the paper is organized as follows: In the next section, we introduce the meta-model EOSs describing nuclear matter and detail the calculations of $\Delta I/I$ and NS crust--core transition properties. Section III discusses symmetry energy effects on the crustal fraction of the NS moment of inertia. Finally, we summarize our findings in Section~IV.

\section{Theoretical~Framework}
\unskip

\subsection{Explicitly Isospin-Dependent Meta-Model EOS for Super-Dense Neutron-Rich Nuclear~Matter}

A meta-model is a model of models. In~the present work, we assume the neutron star matter is made of neutron-rich nuclear matter ($npe\mu$) at $\beta$-equilibrium described by a meta-model constructed in ref.~\citep{Zhang18}.  The~energy density for $npe\mu$ matter at $\beta$-equilibrium can be calculated from
\begin{equation}\label{lepton-density}
  \varepsilon(\rho, \delta)=\rho [E(\rho,\delta)+M_N]+\varepsilon_l(\rho, \delta),
\end{equation}
where the first term donates the energy density of nucleons and $\varepsilon_l(\rho, \delta)$ denotes the energy density for leptons (electrons and muons) that can be calculated from, e.g.,~the
noninteracting Fermi gas model \citep{Oppenheimer39}. $\delta=(\rho_n+\rho_p)/\rho$ is the isospin asymmetry, $M_N$ represents the average nucleon mass of $938$ MeV, and~$E(\rho,\delta)$ is the energy per nucleon of asymmetric nuclear matter, which is a parabolic function of $\delta$ \citep{Bombaci91}:
\begin{equation}\label{Erho}
E(\rho,\delta)=E_0(\rho)+E_{\rm{sym}}(\rho)\cdot\delta ^{2}+\mathcal{O}(\delta^4),
\end{equation}
where $E_0(\rho)$ is the energy per nucleon of symmetric nuclear matter (SNM) and $E_{\rm{sym}}(\rho)$ is nuclear symmetry energy at density $\rho$. Once the energy density is determined, the~baryon densities $\rho_i$ of particle $i$ can be obtained by solving the $\beta$-equilibrium condition $\mu_n-\mu_p=\mu_e=\mu_\mu\approx4\delta E_{\rm{sym}}(\rho)$ where $\mu_i=\partial\varepsilon(\rho,\delta)/\partial\rho_i$ and charge neutrality condition $\rho_p=\rho_e+\rho_\mu$. The~pressure of neutron-rich nuclear matter can be calculated from
\begin{equation}\label{pressure}
P(\rho)=\rho^2\frac{{\rm d}\varepsilon(\rho,\delta(\rho))/\rho}{{\rm d}\rho}.
\end{equation}

Then, an EOS is obtained. To~generate a series of EOSs with continuously variable parameters, we parameterize the $E_0(\rho)$ and $E_{\rm{sym}}(\rho)$ as
\begin{equation}\label{E0-taylor}
  E_{0}(\rho)=E_0(\rho_0)+\frac{K_0}{2}(\frac{\rho-\rho_0}{3\rho_0})^2+\frac{J_0}{6}(\frac{\rho-\rho_0}{3\rho_0})^3,
\end{equation}
\begin{eqnarray}\label{Esym-taylor}
    E_{\rm{sym}}(\rho)&=&E_{\rm{sym}}(\rho_0)+L(\frac{\rho-\rho_0}{3\rho_0})\nonumber\\
    &+&\frac{K_{\rm{sym}}}{2}(\frac{\rho-\rho_0}{3\rho_0})^2
  +\frac{J_{\rm{sym}}}{6}(\frac{\rho-\rho_0}{3\rho_0})^3.
\end{eqnarray}

We emphasize that the above parameterizations are different from the Taylor expansions. Our meta-model EOS constructed for neutron stars contains physics about the compositions of neutron star matter. In~particular, the~proton fraction is determined consistently by the symmetry energy, which plays the most important role in determining the radii of neutron stars. More details about this model can be found in our previous \mbox{publications \citep{Zhang19a,Zhang19b,Zhang2020,Zhang2020b,Zhang2021,Zhang22,Zhang24,Xie24}}.

Based on the accumulations of terrestrial nuclear experiments and astrophysical observations in past decades, the~binding energy $E_0(\rho_0)$ and incompressibility $K_0$ of SNM at the saturation density have been tightly constrained to $E_0(\rho_0)=-15.9\pm0.4$ MeV and $K_0=240\pm20$ MeV \citep{Garg18,Shlomo06}. Based on surveys of 53 analyses of different kinds of terrestrial and
astrophysical data available up to 2016 October, the~magnitude $E_{\rm sym}(\rho_0)$ and slope $L$ of symmetry energy at $\rho_0$ are constrained to $E_{\rm sym}(\rho_0)=31.7\pm3.2$ MeV and \mbox{$L=58.7\pm28.1$ MeV \citep{Li13,Oertel17}}, respectively. For~the parameters characterizing the medium- or high-density behavior of neutron-rich nuclear matter, the~curvature of the symmetry energy is around $K_{\rm sym}=-100\pm100$ MeV \citep{Li21,Grams22,Margueron17b,Mondal17,Somasundaram21}, while the skewness of the SNM EOS is constrained to $J_0=-190\pm100$ MeV \citep{Zhang19,Xie19,Xie20} based on terrestrial experiments and astrophysical observations. Few constraints on $J_{\rm sym}$ have been obtained so far, and~it is only very roughly known to be around $-200<J_{\rm sym}<800$ MeV \citep{Cai17,Zhang17,Tews17}. However, as~a small $J_{\rm sym}$ leads to low-mass neutron stars, especially when $J_{\rm sym}$ is negative, we choose the lower boundary of $J_{\rm sym}$ to be 200 MeV in the present work. We emphasize that the above parameters can be varied independently within the uncertain ranges given above. In~fact, some of the uncertainty ranges are obtained from the marginalized posterior probability distribution functions (PDFs) of individual parameters in Bayesian analyses of NS observables. Nevertheless, once new observables or physics conditions are considered, correlations among the updated posterior EOS parameters may be introduced. In~this work, however, all EOS parameters should be considered independent in the current uncertainty ranges given~above.

Compared with EOSs from microscopic and/or phenomenological nuclear many-body theories, as~well as piecewise polytropes or spectrum functions, here, we want to emphasize the following aspects of the meta-model and justifications for using it. First of all, for~microscopic and/or phenomenological nuclear many-body theories of neutron star matter, though~numerous fundamental physical details can be included, it is challenging to isolate the individual effects of each parameter on the properties of neutron stars as they typically exhibit correlations. On~the contrary, all parameters of our meta-model can be varied independently within their presently known uncertainties. In addition, the~meta-model can generate different EOSs that can mimic essentially all existing EOSs predicted by basically all microscopic and/or phenomenological theories by varying the meta-model EOS parameters. Secondly, we emphasize here that our model is different from the Taylor expansions where the convergence problem arises for densities larger than about 1.5$\rho_0$. {The parameters given by Equations~(\ref{E0-taylor}) and (\ref{Esym-taylor}) coincide with the Taylor expansion coefficients of the EOS around $\rho_0$, but~their high-density interpretation is tied to the specific functional form of the parameterization employed here. Specifically, the~higher-order terms in these equations are not merely coefficients of a Taylor series but are tied to the analytical structure of the chosen parameterization.} As parameterizations, mathematically, they can be valid at any density without the convergence issue even at $(2-3) \rho_0$ and beyond. Their parameters can be extracted from astrophysical observations or terrestrial experiments. On~the other hand, when approaching the saturation density,~Equations~(\ref{E0-taylor}) and (\ref{Esym-taylor}) are exactly the Taylor expansions up to the third order. {The same notation has been consistently adopted in our prior works and more detailed demonstrations can be found in refs. \citep{Zhang18,Zhang19a,Zhang24}.}

At densities below the crust--core transition point, we choose the Negele--Vautherin (NV) EOS \citep{Negele73} for the inner crust and the Baym--Pethick--Sutherland (BPS) EOS \citep{Baym71b} for the outer crust. Once the parameters in Equations~(\ref{E0-taylor}) and (\ref{Esym-taylor}) are fixed and the transition density is calculated, a~unique EOS is obtained. Then, NS properties including masses and radii can be calculated by solving the TOV equations \citep{Tolman34,Oppenheimer39}.

\subsection{Crustal Fraction of Neutron Star Moment of~Inertia}

In the present work, as~we want to check how the observation of the crustal fraction of moment of inertia $\Delta I/I$ can constrain the parameters shown in Equations~(\ref{E0-taylor}) and (\ref{Esym-taylor}), we calculate the $\Delta I/I$ using the following formalism, given in ref.~\citep{Link99}:
\begin{eqnarray}\label{DIIeq}
    \frac{\Delta I}{I}&=&\frac{28\pi P_tR^3}{3Mc^2}\frac{(1-1.67\xi-0.6\xi^2)}{\xi}\nonumber\\
    &\times&[1+\frac{2P_t}{\rho_tm_bc^2}\frac{1+5\xi-14\xi^2}{\xi^2}^{-1}],
\end{eqnarray}
where $\xi=GM/Rc^2$ is the dimensionless compactness, and~$m_b=930$ MeV/c$^2$ is the mass of $^{12}$C/12 or $^{56}$Fe/56 \citep{Lattimer00, Lattimer01}.

As shown in the above equation, the~crustal fraction of the moment of inertia of the neutron star is quite sensitive to the crust--core transition density $\rho_t$ and the pressure $P_t$ there. The~crust--core transition density $\rho_t$ can be found by examining when the incompressibility of NS matter in the uniform core vanishes \citep{Kubis04,Kubis07,Lattimer07}:
\begin{eqnarray}\label{Kmu}
    K_\mu&=&\rho^2\frac{{\rm d^2}E_0}{{\rm d}\rho^2}+2\rho\frac{{\rm d}E_0}{{\rm d}\rho}+\delta^2\nonumber\\
    &\times&[\rho^2\frac{{\rm d^2}E_{\rm sym}}{{\rm d}\rho^2}+2\rho\frac{{\rm d}E_{\rm sym}}{{\rm d}\rho}-2E_{\rm sym}^{-1}(\rho\frac{{\rm d}E_0}{{\rm d}\rho})^2]=0.
\end{eqnarray}

Once $K_\mu$ becomes negative, the~speed of sound becomes imaginary. It indicates the onset of the mechanical instability (spinodal decomposition of matter) in the core, leading to the formation of
clusters making up the crust.
This method has been used extensively in the literature to locate the crust--core transition point using various EOSs (see, \mbox{e.g.,~\citep{Xu09,Ducoin11,Piekarewicz14,Routray16}}). As~the $E_{\rm sym}$ and $E_0$ appear explicitly in the above equation, it is thus necessary and interesting to explore whether $E_{\rm sym}$ and $E_0$ have significant effects on the crustal fraction of the NS moment of inertia \citep{Worley,Xu09b}. The~flexibility of the meta-model and the very diverse EOSs generated with it
enable us to explore possible effects of $E_{\rm sym}$ and $E_0$ on $\Delta I/I$ more easily and broadly. In~particular, we focus on the effects of the high-order coefficients of nuclear symmetry energy ($K_{\rm sym}$ and $J_{\rm sym}$) that have not been studied yet in the literature. Since the
crust--core transition density and pressure are determined by the curvature of NS matter in Equation~(\ref{Kmu}), the~$K_{\rm sym}$ and $J_{\rm sym}$ are expected to be important for determining the $\Delta I/I$.

\section{Results and~Discussions}
\unskip

\subsection{Effects of Nuclear EOS Parameters on the Crustal Fraction of the NS Moment of~Inertia}

\begin{figure*}[ht]

  \includegraphics[width=15cm]{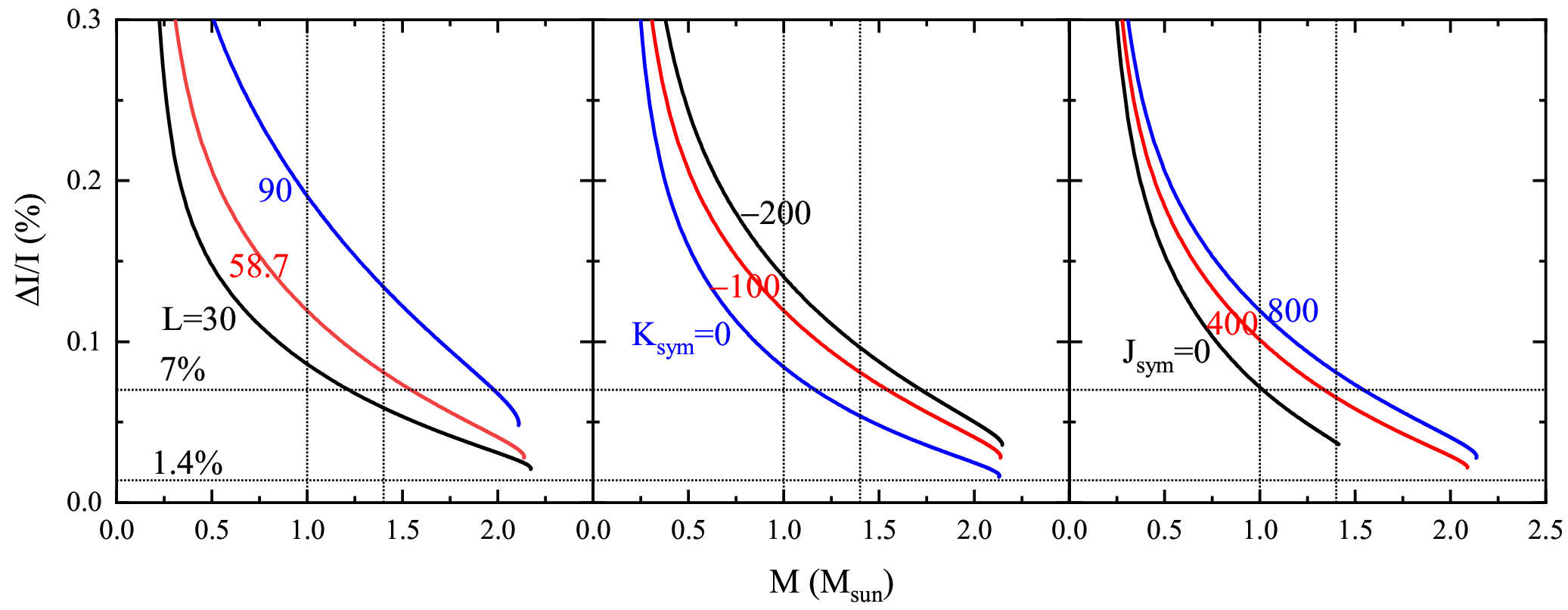}
  \caption{The effects of the slope $L$ (left plot), culture $K_{\rm sym}$ (middle plot), and~skewness $J_{\rm sym}$ (right plot) of symmetry energy on the crustal fraction of NS moment of inertia $\Delta I/I$. The~horizontal dotted lines represent the constraints of 1.4\% and 7\%, while the vertical dotted lines correspond to neutron star masses of 1.0 and 1.4 M$_\odot$. }\label{DIIEsym}
\end{figure*}

Effects of the slope $L$ (left plot), curvature $K_{\rm sym}$ (middle plot), and~skewness $J_{\rm sym}$ (right plot) of nuclear symmetry energy on the crustal fraction of NS moment of inertia $\Delta I/I$ are illustrated in Figure~\ref{DIIEsym}. The~horizontal dotted lines represent the constraints of 1.4\% and 7\%, while the vertical dotted lines correspond to neutron star masses of 1.0 and 1.4 M$_\odot$. Other parameters are kept at the most probable values currently known, as previously mentioned. From~the figure, it is evident that the symmetry energy parameters exert minimal influence on the NS maximum mass reached, with~the exception of $J_{\rm sym}=0$, which results in a very soft EOS at high densities. Additionally, all symmetry energy parameters significantly affect the crustal fraction of the NS moment of inertia. More specifically, as~$L$ increases, the~$\Delta I/I$ consistently increases for a given NS mass. The~condition $\Delta I/I>1.4\%$ is easily satisfied, whereas $\Delta I/I>7\%$ is only met when $M<1.21$ M$_\odot$ for $L=30$ MeV and $M<1.98$ M$_\odot$ for $L=90$ MeV, respectively. This indicates that the constraint of $\Delta I/I>7\%$ can exclude $L=30$ MeV for neutron stars with masses larger than $1.21$ M$_\odot$ while favoring larger values of $L$ that allow for a broader mass range to satisfy the $\Delta I/I>7\%$
constraint.

A similar trend is observed for $K_{\rm sym}$ and $J_{\rm sym}$. However, an~increase in $K_{\rm sym}$ leads to a decrease in $\Delta I/I$. This is due to the fact that a higher $K_{\rm sym}$ results in a lower crust--core transition density, particularly affecting the transition pressure. On~the other hand, an~increase in $L$ or $J_{\rm sym}$ raises the crust--core transition density, as~shown in \mbox{Figure~3 of ref.~\citep{Zhang18}}. It is important to note that, since the symmetry energy at saturation density is well constrained, the~effects of $E_{\rm sym}(\rho_0)$ are not significant and are thus not depicted here. Consequently, the~competition among the symmetry energy parameters ultimately determines the value of $\Delta I/I$ and its mass~dependence.

\begin{figure}[ht]
\includegraphics[width=8 cm]{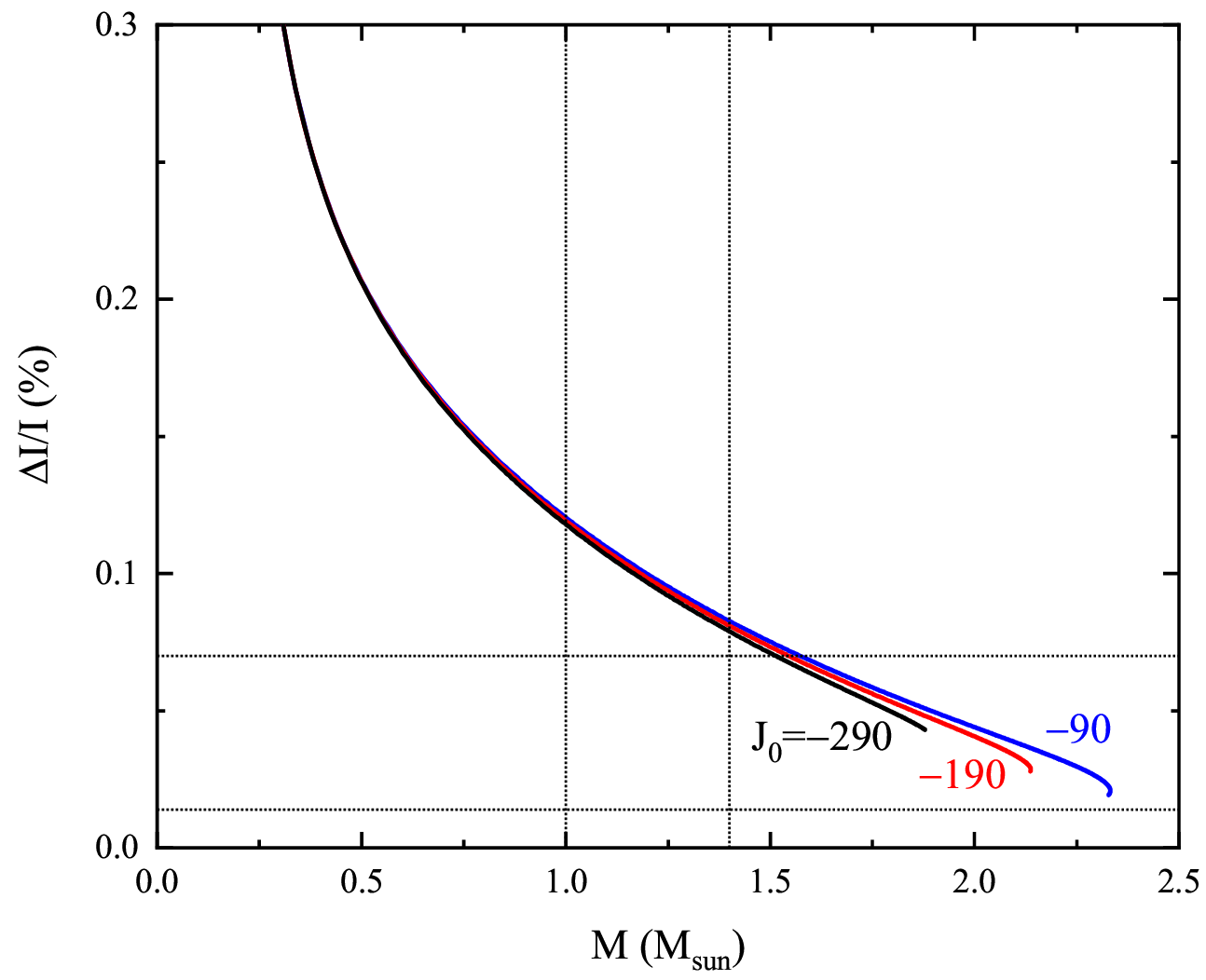}
\caption{\label{DIIE0} Same as Figure~\ref{DIIEsym} but for the skewness $J_0$ of~SNM.}
\end{figure}

In addition to the effects of nuclear symmetry energy, effects of the skewness $J_0$ of SNM on the crustal fraction of NS moment of inertia $\Delta I/I$ are illustrated in Figure~\ref{DIIE0}. It was seen that $J_0$ has minimal impact on $\Delta I/I$, leading us to fix $J_0=-190$ MeV in the subsequent discussions regarding $\Delta I/I$. However, $J_0$ does significantly influence the maximum mass of neutron stars, which increases from 1.88 to 2.33 M$_\odot$ as $J_0$ rises from $-290$ MeV to \mbox{$-90$ MeV}. Similar findings have been reported in our previous publications (e.g., refs.~\citep{Zhang18, Zhang19a, Zhang19b}). Likewise, effects of the incompressibility $K_0$ of SNM are not shown here as~it is well constrained and has little effect on the $\Delta I/I$.

Based on the discussions above, we conclude that, while the symmetry energy can significantly affect the moment of inertia, the~EOS of SNM has minimal impact. Thus, we will focus on the effects of symmetry energy in the following~subsection.

\subsection{Constraints on Symmetry Energy Parameters Imposed by Crustal Fraction of NS Moment of~Inertia}

Currently, the~lightest mass of observed NSs is around 1.0 M$_\odot$, while most NSs have masses around 1.4 M$_\odot$, {and the maximum observed mass is around 2.0 M$_\odot$. For~the integrity of our following discussions, the~NSs with masses of 1.0 M$_\odot$, 1.4 M$_\odot$, and~2.0 M$_\odot$} are included as~examples.

\begin{figure*}[ht]
\includegraphics[width=13.8 cm]{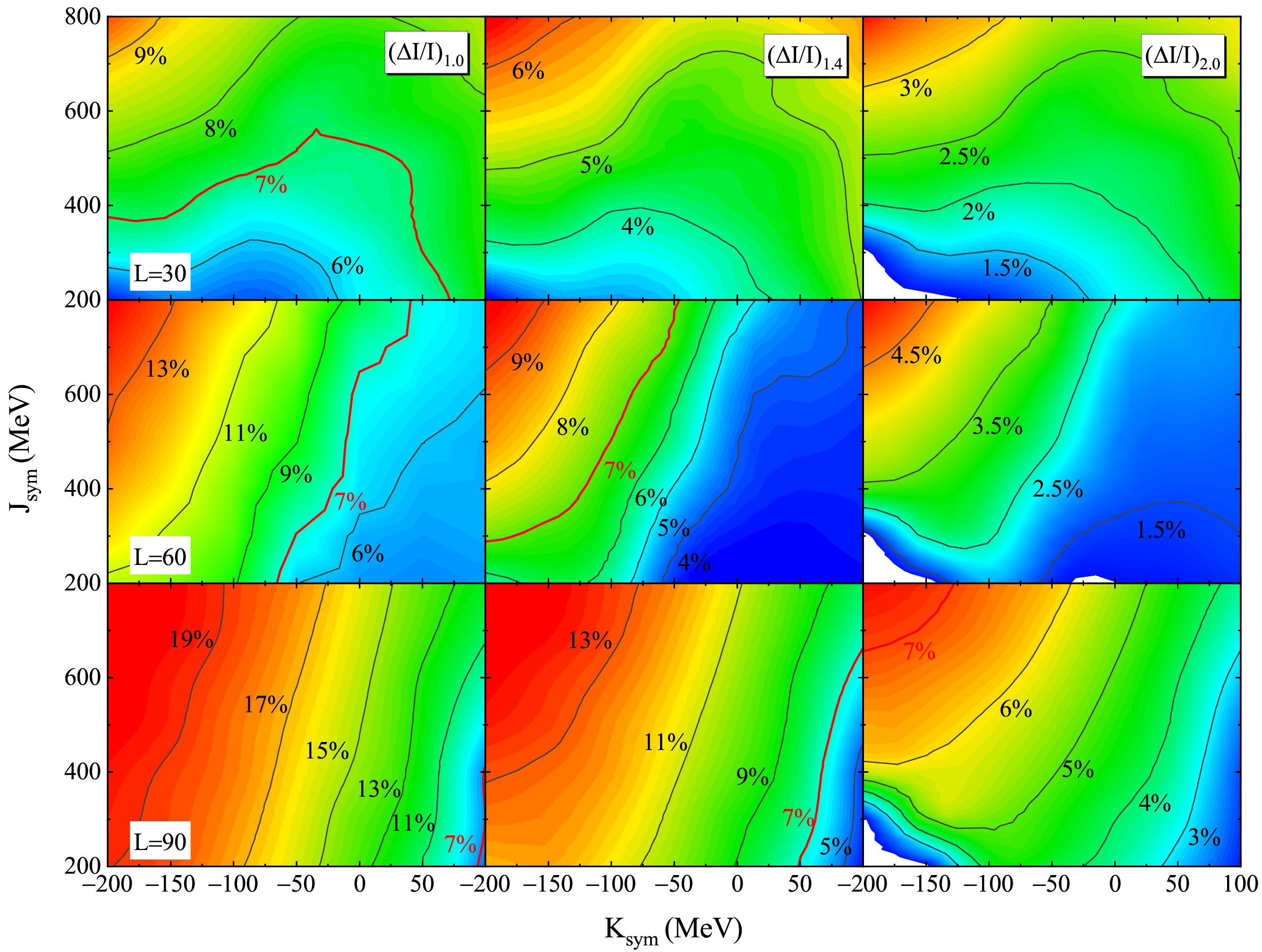}
\caption{\label{DII}{The counter of the crustal fraction of NS moment of inertia $\Delta I/I$ in the plane of $K_{\rm sym}$ and $J_{\rm sym}$ for the neutron stars with 1.0 M$_\odot$ (left plot), 1.4 M$_\odot$ (middle plot), and~2.0  M$_\odot$ (right plot) for slope of symmetry energy $L=30$ (upper plots), $60$ (middle plots), and~$90$ MeV (lower plots), respectively. The~red lines in the left and middle plots indicate the constraint of $\Delta I/I=7\%$.}}
\end{figure*}

Fixing the lower mass limit of neutron stars at 1.0 M$_\odot$, the~left plots of Figure~\ref{DII} present the contour of the crustal fraction of NS moment of inertia $\Delta I/I$ in the $K_{\rm sym}$ versus $J_{\rm sym}$ plane for the slope of symmetry energy $L=30$ (upper plot), $60$ (middle plot), and~$90$ MeV (lower plot), respectively. The~red lines indicate the constraint of $\Delta I/I=7\%$. It is apparent that $\Delta I/I\geq 7\%$ or $1.4\%$ can be satisfied for all values of $L$ considered. The~maximum $\Delta I/I$ appears at $K_{\rm sym}=-200$ and $J_{\rm sym}=800$ MeV, as~previously shown in Figure~\ref{DIIEsym}. $\Delta I/I$ increases with increasing $J_{\rm sym}$ but decreases with $K_{\rm sym}$. Additionally, the~allowed range expands as
$L$ increases.

For $L=30$ MeV, the~area below the red line is excluded. Although~the individual uncertainties in $K_{\rm sym}$ and $J_{\rm sym}$ cannot be narrowed further compared to their initial uncertainties, the~combinations of these parameters are significantly constrained. In~other words, nearly half of the parameter plane is excluded, particularly when both $K_{\rm sym}$ and $J_{\rm sym}$ are small. Furthermore, the~relationship between $K_{\rm sym}$ and $J_{\rm sym}$ for constant $\Delta I/I$ is not monotonic. As~$K_{\rm sym}$ increases, the~curves for fixed $\Delta I/I$ initially rise and then fall. At~$K_{\rm sym}=-200$ MeV, $J_{\rm sym}$ must exceed $294$ MeV, which can be explained by examining the transition density and pressure, as~discussed~below.

For $L=60$ MeV, the~relationship between $K_{\rm sym}$ and $J_{\rm sym}$ for constant $\Delta I/I$ becomes monotonic, and~the lower-right region of the plane is excluded. The~upper limit of $K_{\rm sym}$ is constrained to be $41$ MeV, while $J_{\rm sym}$ remains unconstrained for $L=60$ MeV. Thus, the~crustal fraction of the NS moment of inertia can help constrain the symmetry energy at medium densities characterized by $K_{\rm sym}$, rather than at high densities, for~larger values of $L$. When $L$ increases to $90$ MeV, only a negligible range is excluded to satisfy $\Delta I/I\geq7\%$ in the lower-right corner. Given that the upper limit of $K_{\rm sym}$ is constrained to be about 0, symmetry energies with larger values of $L$ cannot be constrained at any density by the $\Delta I/I\geq7\%$ constraint.

Although a neutron star with 1.0 M$_\odot$ can achieve $\Delta I/I\geq7\%$ for any value of $L$, we find that (1) a larger $J_{\rm sym}$ is favored for $L=30$ MeV; (2) a smaller $K_{\rm sym}$ is favored for $L=60$ MeV, while no constraints can be drawn for $J_{\rm sym}$; and (3) no constraints can be extracted for $K_{\rm sym}$ and $J_{\rm sym}$ for $L=90$ MeV. Therefore, the~symmetry energy should be treated with caution when calculating the moment of inertia of neutron stars. Additionally, the~allowed parameter plane enlarges with $L$, indicating that a stiffer symmetry energy around saturation density, characterized by $L$, is more likely to satisfy the constraints imposed by the crustal fraction of the NS moment of~inertia.

For a canonical NS with $M=1.4$ M$_\odot$, the~contours of the crustal fraction of NS moment of inertia $\Delta I/I$ in the $K_{\rm sym}$ versus $J_{\rm sym}$ plane, with~$L=30$ (upper plot), $60$ (middle plot), and~$90$ MeV (lower plot), respectively, are presented in the middle plots of Figure~\ref{DII}. When compared to the left plots of Figure~\ref{DII}, similar features can be observed. However, the~parameter plane is clearly more tightly constrained for any value of $L$ as the crustal fraction of the NS moment of inertia decreases with increasing mass. With~$L=30$ MeV, no parameter combinations can satisfy the constraint $\Delta I/I\geq7\%$, indicating that this constraint favors a symmetry energy with a larger slope $L$. For~instance, at~$L=60$ MeV, we obtain tighter constraints on $K_{\rm sym}$ and $J_{\rm sym}$ compared to a neutron star with \mbox{$M=1.0$ M$_\odot$}. The~lower limit of $J_{\rm sym}$ is constrained to be $289$ MeV, while the upper limit of $K_{\rm sym}$ is constrained to be $-46$ MeV. This constrained value of $K_{\rm sym}$ is consistent with findings of previous studies that suggest $K_{\rm sym}=-100\pm100$ MeV \citep{Li21, Grams22, Margueron17b, Mondal17, Somasundaram21}. These constraints on $K_{\rm sym}$ and $J_{\rm sym}$ may indicate a relatively stiff symmetry energy at high densities. For \mbox{$L=90$ MeV}, the~lower-right corner of the plane is excluded. However, if~the upper limit of $K_{\rm sym}$ is set to 0, the~crustal fraction of the NS moment of inertia still cannot effectively constrain the symmetry~energy.

{For the most massive observed NS with $M\approx2.0$ M$_\odot$, the~contours of $\Delta I/I$ in the $K_{\rm sym}$ versus $J_{\rm sym}$ plane are presented in the right plots of Figure~\ref{DII}. The~white regions indicate the excluded parameter sets that cannot support an NS with $M=2.0$ M$_\odot$. It is clear that the constraint of $\Delta I/I=7\%$ can only be satisfied within a narrow range for $L=90$ MeV.}

In short, the~constraint $\Delta I/I\geq7\%$ does not favor symmetry energies with small slopes (such as $L=30$ MeV), and~tighter constraints on symmetry energy can be extracted for neutron stars with masses around 1.4 {or 2.0 M$_\odot$}. If~the neutron star has a significantly larger mass, it is anticipated that only symmetry energies with a considerably larger slope can satisfy this~constraint.

From Equation~(\ref{Kmu}), it is evident that the transition pressure $P_t$ plays the most important role in determining the crustal fraction of the moment of inertia, alongside the mass $M$ and radius $R$ of an NS. Notably, the~transition density $\rho_t$ merely appears in the correction term and thus contributes little to the moment of inertia. Furthermore, lower values of $P_t$ result in a decreased crustal moment of inertia, thereby imposing stricter constraints on the EOS or the symmetry energy $E_{\rm sym}$. {In the thermodynamical approach given in Equation~(\ref{Kmu}), the~crust--core transition pressure can be approximated as follows \citep{Lattimer00,Lattimer01,Lattimer07}:
\begin{eqnarray}
   P_t&=& \frac{K_0}{9}\frac{\rho_t^2}{\rho_0}\left(\frac{\rho_t}{\rho_0}-1\right) \\ \nonumber 
   &+&  \rho_t\delta_t\left[\frac{1-\delta_t}{2}E_{\rm sym}(\rho_t)+\left(\rho\frac{{\rm d}E_{\rm sym}(\rho)}{{\rm d}\rho}\right)_{\rho_t}\delta_t\right].\nonumber 
\end{eqnarray}

The above equation shows clearly that the relation between $P_t$ and $\rho_t$ is complicated. The~Pearson correlation coefficient between them is calculated by generating about 39,000~EOSs where all parameters vary within their uncertainties. The~resulting Pearson coefficient of 0.92631 indicates a strong and approximately linear correlation between them. Thus, the~correlations between either $P_t$ or $\rho_t$ and the EOS parameters exhibit a strong similarity.
}
The behavior of the constant crustal fraction of the NS moment of inertia can be explained through the core--crust transition density and pressure. In \mbox{Figure~3 of ref.~\citep{Zhang18}}, we plotted the contours of transition density $\rho_t$ and transition pressure $P_t$ in the $L$ and $K_{\rm sym}$ plane with fixed $J_{\rm sym}$. We found that, for~constant $\rho_t$, $K_{\rm sym}$ exhibits a monotonic relationship with $L$ when $K_{\rm sym}<-100$ MeV. However, this relationship becomes more complex for larger values of $K_{\rm sym}$.

\begin{figure*}[ht]
\includegraphics[width=13 cm]{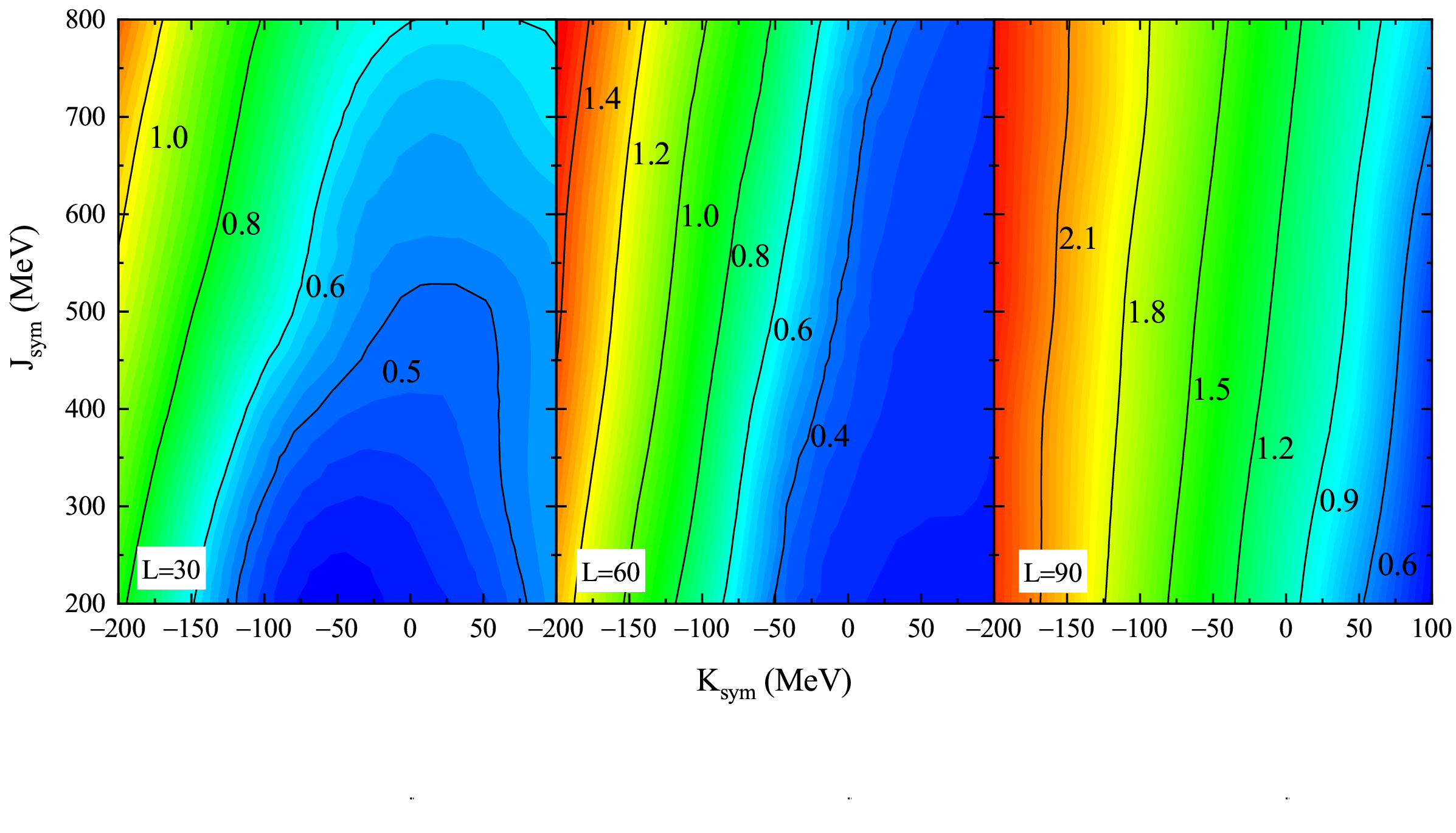}
\vspace{-1 cm}
\caption{\label{Pt} The counter of transition pressure $P_t$ for slope of symmetry energy $L=30$ (left plot), $60$~(middle plot), and~$90$ MeV (right plot).}
\end{figure*}

To illustrate the effects of $J_{\rm sym}$ on the transition pressure, we present the contours of transition pressure $P_t$ in the $K_{\rm sym}$ versus $J_{\rm sym}$ plane in Figure~\ref{Pt}. It is evident that the trends for a constant transition pressure are quite similar to those for the constant moment of inertia shown in Figure~\ref{DII}. Notably, the~consistency decreases with increasing neutron star mass, which is understandable since the crust of a neutron star is thicker and constitutes a larger mass percentage in lower-mass stars. Consequently, the~core--crust transition properties have a more pronounced effect on low-mass neutron~stars.

\begin{figure*}[ht]

  \includegraphics[width=13cm]{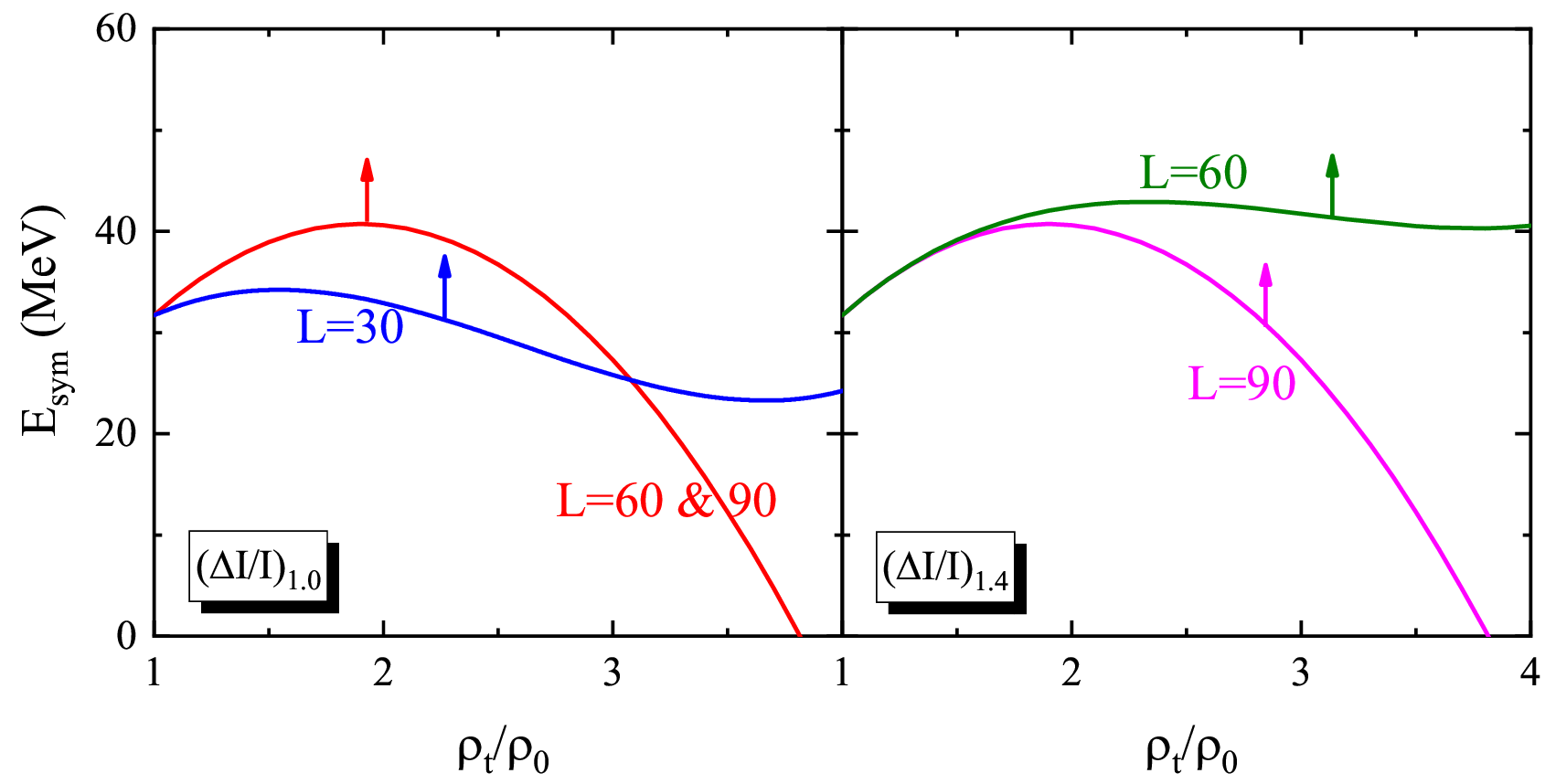}
  \caption{The lower boundaries of symmetry energy constrained for different slope $L$ of symmetry energy for the neutron stars with 1.0 M$_\odot$ (left plot) and 1.4 M$_\odot$ (right plot), respectively.}\label{Esymlm}
\end{figure*}

As $L$, $K_{\rm sym}$, and~$J_{\rm sym}$ are the three parameters characterizing the density dependence of nuclear symmetry energy, the~constraints on these parameters can be translated into constraints on the symmetry energy itself. The~lower boundaries of the symmetry energy for different slopes $L$ of the symmetry energy, corresponding to neutron stars with 1.0 M$_\odot$ (left plot) and 1.4 M$_\odot$ (right plot), are illustrated in Figure~\ref{Esymlm}. It is important to note that the lower limit of $E_{\rm sym}$ is determined by the left bottom corner, specifically, $K_{\rm sym}=-200$ MeV and $J_{\rm sym}=200$ MeV within the parameter uncertainties, assuming we do not consider the constraints from the crustal fraction of the NS moment of inertia. With~$L=60$ and $90$ MeV in the case of the neutron star with 1.0 M$_\odot$, and~for $L=90$ MeV for the neutron star with 1.4~M$_\odot$, the~lower limit of the symmetry energy remains the same, as~the left bottom point is not excluded (as shown in Figure~\ref{DII}). As~the transition pressure remains almost constant for most combinations of $K_{\rm sym}$ and $J_{\rm sym}$ in the left plot of Figure~\ref{Pt} for $L=30$ MeV, a larger $J_{\rm sym}$ is needed to satisfy $\Delta I/I\geq7\%$ for the neutron star with 1.0 M$_\odot$ and $J_{\rm sym}$ cannot be smaller than $294$ MeV (a larger $K_{\rm sym}$ leads to a smaller $\Delta I/I$). Consequently, the~lower limit of symmetry energy along the red line in the upper plot of of left plot of Figure~\ref{DII} can be extracted, and~its lower limit is represented by the blue line in the left plot of Figure~\ref{Esymlm}.

Since no combination of parameters can satisfy $\Delta I/I\geq7\%$ for neutron stars with 1.4~M$_\odot$ at $L=30$ MeV, only the lines for $L=60$ MeV and $90$ MeV are plotted in the right plot of Figure~\ref{Esymlm}. Focusing on the blue and green lines, it is clear that the symmetry energy remains nearly constant with increasing density, even for densities $\rho>3\rho_0$. It is worth noting that the symmetry energy can only be constrained for approximately $\rho<3\rho_0$ based on the current observations of NSs by NICER \citep{Miller21, Riley21} and LIGO and Virgo \citep{LIGO18} Collaborations, as~discussed in our previous study \citep{Zhang2021}. The~present results suggest that the crustal fraction of the NS moment of inertia has the potential to constrain the lower limit of the symmetry energy at densities exceeding $3\rho_0$, particularly when $L$ is constrained to values less than $60$~MeV.

\section{Summary and~Conclusions}
In summary, using an isospin-dependent NS meta-model EOS, we investigated the implications of the crustal fraction of NS moment of inertia $\Delta I/I$ for NSs with 1.0 M$_\odot$ or 1.4 M$_\odot$ on the density dependence of nuclear symmetry energy $E_{\rm sym}(\rho)$. We found that an increase in the slope $L$ and skewness $J_{\rm sym}$ of symmetry energy results in a larger $\Delta I/I$, whereas an increase in the curvature $K_{\rm sym}$ leads to a reduction in $\Delta I/I$. On~the other hand, the SNM EOS was found to have little influence on the $\Delta I/I$.
Our findings highlight the significant role of the symmetry energy parameters in determining the crustal fraction of the NS moment of inertia. Moreover, we found that a larger slope $L$ is favored to satisfy the $\Delta I/I\geq7\%$ constraint, with~tighter constraints emerging for neutron stars around 1.4 M$_\odot$. Furthermore, we extracted the lower boundaries for $E_{\rm sym}(\rho)$ from the constraint $\Delta I/I\geq7\%$ for different $L$ values. Our results indicate that the crustal fraction of the NS moment of inertia has the potential to set a lower limit of symmetry energy at densities exceeding $3\rho_0$, particularly when $L$ can be constrained to values less than $60$ MeV, reflecting the complex interplay between the core--crust transition properties and the underlying nuclear interactions. These findings emphasize the potential of using the crustal fraction of the NS moment of inertia to refine our understanding of NS physics and the properties of supradense neutron-rich nuclear~matter.

Of course, our work has limitations and caveats. First, the~crustal moment of inertia is highly sensitive to the crust--core transition density. It is well known that employing different crust models can introduce uncertainty into the radius calculations. However, a quantitative estimate of the associated uncertainty in the predicted NS radius is model dependent, ranging from about 0.1 to 0.7 km (Table I of ref. \citep{Fortin}), while its quantitative effects on the crustal moment of inertia are still unclear. Unified EOSs for both the crust and the core from the same microscopic and/or phenomenological nuclear many-body theories (\mbox{e.g., refs.~\citep{Carreau19, Davis24})} could describe the core--crust transition properties more consistently. While our meta-model can mimic some of these uncertainties by varying the EOS parameters as discussed above, investigating the associated microscopic physics in and around the crust--core transition is beyond the ability of our meta-model.
Similarly, when deriving the crustal moment of inertia from pulsar observations, the~complex dynamics of nucleon superfluidity, which is closely related to the nuclear force, must be considered. Here, we used the observational constraint on the crustal fraction of the neutron star moment of inertia regardless of how it is extracted. Within~the framework of our meta-model, we cannot say anything about the microphysics occurring around the interface between the core and the crust of neutron stars. Thirdly, the~thermodynamical method (via Equation~(\ref{Kmu})) suggests larger values of $P_t$ or $\rho_t$ (and, consequently, larger values of $\Delta I/I$) compared to the dynamical method, as shown in refs. \citep{Xu09, Tsaloukidis19}. Therefore, tighter constraints may be obtained if the dynamical method is adopted to calculate the $P_t$ or $\rho_t$. Nevertheless, our main qualitative conclusions are not expected to be affected by the above~uncertainties.

\vspace{6pt}

\noindent{\bf Acknowledgments}\\
We would like to thank Armen Sedrakian and the organizing committee of the
Modern Physics of Compact Stars conference series for providing a productive and inspiring platform where some results of the present work were presented and discussed. B.A.L. is supported in part by the U.S. Department of Energy, Office of Science, under~award number DE-SC0013702, and the~CUSTIPEN (China--U.S. Theory Institute for Physics with Exotic Nuclei) under the U.S. Department of Energy grant no. DE-SC0009971. N.B.Z. is supported in part by the National Natural Science Foundation of China under grant no. 12375120, the~Zhishan Young Scholar of Southeast University under grant no. 2242024RCB0013, and the Start-Up Research Fund of Southeast University under grant no. RF1028623060.


\begin{thebibliography}{999}
\bibitem[Xu(2010)]{Xu10} Xu, C.; Li, B.A. Understanding the major uncertainties in the nuclear symmetry energy at suprasaturation densities. {\em Phys. Rev. C} {\bf 2010}, {\em 81}, 064612.

\bibitem{Plamen}Krastev, P.G.; Li, B.A.
Imprints of the nuclear symmetry energy on the tidal deformability of neutron stars. {\em
J. Phys. G} {\bf 2019}, {\em 46},~074001.

\bibitem[Lattimer(2000)]{Lattimer00} Lattimer J.M.; Prakash M. Nuclear matter and its role in supernovae, neutron stars and compact object binary mergers. {\em Phys. Rep.} {\bf 2000}, {\em 333}, 121--146.
\bibitem[Lattimer(2001)]{Lattimer01} Lattimer J.M.; Prakash M. Neutron star structure and the equation of state. {\em Astrophys. J.} {\bf 2001}, {\em 550}, 426--442.
\bibitem[Danielewicz(2002)] {Danielewicz02} Danielewicz, P.; Lacey, R.; Lynch, W.G. Determination of the Equation of State of Dense Matter. {\em Science} {\bf 2002}, {\em 298}, 1592--1596.

\bibitem{Ebook}
Li, B.A.; Ramos, A.; Verde, G.; Vidana, I.
Topical issue on nuclear symmetry energy.
{\em Eur. Phys. J. A} {\bf 2014}, {\em 50}, 9.

\bibitem[Lattimer(2016)]{Lattimer16} Lattimer J.M.; Prakash M. The equation of state of hot, dense matter and neutron stars. {\em Phys. Rep.} {\bf 2016}, {\em 621}, 127--164.
\bibitem[Watts(2016)]{Watts16} Watts, A.L.; Andersson, N.; Chakrabarty, D.; Feroci, M.; Hebeler, K.; Israel, G.; Lamb, F.K.; Miller, M.C.; Morsink, S.; \"{O}zel, F.;~et~al. Colloquium: Measuring the neutron star equation of state using x-ray timing. {\em  Rev. Mod. Phys.} {\bf 2016}, {\em 88}, 021001.
\bibitem[Oertel(2017)]{Oertel17} Oertel, M.; Hempel, M.; Kl\"{a}hn, T.; Typel, S. Equations of state for supernovae and compact stars. {\em  Rev. Mod. Phys.} {\bf 2017}, {\em 89}, 015007.
\bibitem[Ozel(2016)]{Ozel16} \"{O}zel, F.; Freire, P. Masses, radii, and the equation of state of neutron stars. {\em  Annu. Rev. Astron. Astrophys.} {\bf 2016}, {\em 54}, 401-440.
\bibitem[Blaschke(2018)]{Blaschke18} Blaschke, D.; Chamel, N. Phases of Dense Matter in Compact Stars. In {\em The Physics and Astrophysics of Neutron Stars}; Rezzolla, L., Pizzochero, P., Jones, D.I., Rea, N., Vida\~{n}a, I., Eds.; Astrophysics and Space Science Library; Springer: Cham, Switzerland, 2018; Volume 457, pp. 337--400.
\bibitem[Cromartie(2019)]{Cromartie19} Cromartie, H.T.; Fonseca, E.; Ransom, S.M.; Demorest, P.B.; Arzoumanian, Z.; Blumer, H.; Brook, P.R.;  DeCesar, M.E.; Dolch, T.; Ellis, J.A.;~et~al. Relativistic Shapiro delay measurements of an extremely massive millisecond pulsar. {\em Nat. Astron.} {\bf 2019}, {\em 4}, 72--76.
\bibitem[Fonseca(2021)]{Fonseca21} Fonseca, E.; Cromartie, H.T.; Pennucci, T.T.; Ray, P.S.; Kirichenko, A.Y.; Ransom, S.M.; Demorest, P.B.;  Stairs, I.H.; Arzoumanian, Z.; Guillemot, L.;~et~al. Refined mass and geometric measurements of the high-mass PSR J0740+6620. {\em Astrophys. J. Lett.} {\bf 2021}, {\em 915},~L12.
\bibitem[Miller(2021)]{Miller21} Miller, M.C.; Lamb, F.K.; Dittmann, A.J.; Bogdanov, S.; Arzoumanian, Z.; Gendreau, K.C.; Guillot, S.; Ho, W.C.G.; Lattimer, J.M.; Loewenstein, M.;~et~al. The radius of PSR J0740+ 6620 from NICER and XMM-Newton data. {\em Astrophys. J. Lett.} {\bf 2021}, {\em 918},~L28.
\bibitem[Riley(2021)]{Riley21} Riley, T.E.; Watts, A.L.; Ray, P.S.; Bogdanov, S.; Guillot, S.; Morsink, S.M.; Bilous, A.V.; Arzoumanian, Z.; Choudhury, D.; Deneva, J.S.;~et~al. A NICER view of the massive pulsar PSR J0740+ 6620 informed by radio timing and XMM-Newton spectroscopy. {\em Astrophys. J. Lett.} {\bf 2021}, {\em 918}, L27.
\bibitem[Salmi(2022)]{Salmi22} Salmi, T.; Vinciguerra, S.; Choudhury, D.; Riley, T. E.; Watts, A. L.; Remillard, R. A.; Ray, P.S.; Bogdanov, S.; Guillot, S.; Arzoumanian, Z.;~et~al. The radius of PSR J0740+ 6620 from NICER with NICER background estimates. {\em Astrophys. J.} {\bf 2022}, {\em 941},~150.
\bibitem[Salmi(2024)]{Salmi24} Salmi, T.; Choudhury, D.; Kini, Y.; Riley, T.E.; Vinciguerra, S.; Watts, A.L.; Wolff, M.T.; Arzoumanian, Z.; Bogdanov, S. The Radius of the High-mass Pulsar PSR J0740+6620 with 3.6 yr of NICER Data. {\em Astrophys. J.} {\bf 2024}, {\em 974}, 294.
\bibitem[Dittmann(2024)]{Dittmann24} Dittmann, A.J.; Miller, M.C.; Lamb, F.K.; Holt, I.; Chirenti, C.; Wolff, M.T.; Bogdanov, S.; Guillot, S.; Ho, W.C.G.; Morsink, S.M.;~et~al. A more precise mMeasurement of the radius of PSR J0740+6620 using updated NICER data. \emph{arXiv} \textbf{2024}, arXiv:2406.14467.
\bibitem[Abbott(2018)]{LIGO18} Abbott, B.P.; Abbott, R.; Abbott, T.D.; Acernese, F.; Ackley, K.; Adams, C.; Adams, T.; Addesso, P.; Adhikari, R.X.; Adya, V.B.;~et~al. GW170817: Measurements of neutron star radii and equation of state. {\em Phys. Rev. Lett.} {\bf 2018}, {\em 121}, 161101.
\bibitem[Baiotti(2019)]{Baiotti19} Baiotti, L. Gravitational waves from neutron star mergers and their
relation to the nuclear equation of state. {\em Prog. Part. Nucl. Phys.} {\bf 2019}, {\em 109}, 103714.
\bibitem[Burgio(2021)]{Burgio21} Burgio, G.F.; Schulze, H.J.; Vida\~{n}a, I.; Wei, J.B. Neutron stars and the nuclear equation of state. {\em Prog. Part. Nucl. Phys.} {\bf 2021}, {\em 120},~103879.
\bibitem[Li(2021)]{Li21} Li, B.A.; Cai, B.J.; Xie, W.J.; Zhang, N.B. Progress in constraining nuclear symmetry energy using neutron star observables since GW170817. {\em Universe} {\bf 2021}, {\em 7}, 182.
\bibitem[Hooker(2015)]{Hooker15} Hooker, J.; Newton, W.G.; Li, B.A. Efficacy of crustal superfluid neutrons in pulsar glitch models. {\em Mon. Not. R. Astron. Soc.} {\bf 2015}, {\em 449}, 3559--3567.
\bibitem[Newton(2015)]{Newton15} Newton, W.G.; Berger, S.; Haskell, B. Observational constraints on neutron star crust-core coupling during glitches. {\em Mon. Not. R. Astron. Soc.} {\bf 2015}, {\em 454}, 4400--4410.
\bibitem[Liu(2018)]{Liu18} Liu, Z.W.; Qian, Z.; Xing, R.Y.; Niu, J.R.; Sun, B.Y. Nuclear fourth-order symmetry energy and its effects on neutron star properties in the relativistic Hartree-Fock theory. {\em Phys. Rev. C} {\bf 2018}, {\em 97}, 025801.
\bibitem[Dutra(2021)]{Dutra21} Dutra, M.; Lenzi, C.H.; de Paula, W.; Louren\c{c}o, O. Neutron star crustal properties from relativistic mean-field models and bulk parameters effects. {\em Euro. Phys. J. A} {\bf 2021}, {\em 57}, 260.
\bibitem[ Parmar(2022)]{Parmar22} Parmar, V.; Das, H.C.; Kumar, A.; Sharma, M.K.; Patra, S.K. Crustal properties of a neutron star within an effective relativistic mean-field model. {\em Phys. Rev. D} {\bf 2022}, {\em 105}, 043017.
\bibitem[Radhakrishnan(1969)]{Radhakrishnan69} Radhakrishnan, V.; Manchester, R.N. Detection of a change of state in the pulsar PSR 0833-45. {\em Nature} {\bf 1969}, {\em 222}, 228--229.
\bibitem[Reichley(1969)]{Reichle69} Reichley, P.E.; Downs, G.S. Observed decrease in the periods of pulsar PSR 0833-45. {\em Nature} {\bf 1969}, {\em 222}, 229--230.
\bibitem[Anderson(1975)]{Anderson75} Anderson, P.W.; Itoh, N. Pulsar glitches and restlessness as a hard superfluidity phenomenon. {\em Nature} {\bf 1969}, {\em 256}, 25--27.
\bibitem[Ruderman(1976)]{Ruderman76} Ruderman, M. Crust-breaking by neutron superfluids and the VELA pulsar glitches. {\em Astrophys. J.} {\bf 1976}, {\em 203}, 213.
\bibitem[Pines(1985)]{Pines85} Pines, D.; Alpar, M.A. Superfluidity in neutron stars. {\em Nature} {\bf 1985}, {\em 316}, 27--32
\bibitem[Link(1999)]{Link99} Link, B.; Epstein, R.I.; Lattimer, J.M. Pulsar constraints on neutron star structure and equation of state. {\em Phys. Rev. Lett.} {\bf 1999}, {\em 83},~3362.
\bibitem[Espinoza(2011)]{Espinoza11} Espinoza, C.M.; Lyne, A.G.; Stappers, B.W.; Kramer, M. A study of 315 glitches in the rotation of 102 pulsars. {\em Mon. Not. R. Astron. Soc.} {\bf 2011}, {\em 414}, 1679--1704.
\bibitem[Andersson(2012)]{Andersson12} Andersson, N.; Glampedakis, K.; Ho, W.C.; Espinoza, C.M. Pulsar glitches: The crust is not enough. {\em Phys. Rev. Lett.} {\bf 2012}, {\em 109},~241103.
\bibitem[Chamel(2012)]{Chamel12} Chamel, N. Neutron conduction in the inner crust of a neutron star in the framework of the band theory of solids. {\em Phys. Rev. C} {\bf 2012}, {\em 85}, 035801.
\bibitem[Fattoyev(2010)]{Fattoyev10} Fattoyev, F.J.; Piekarewicz, J. Sensitivity of the moment of inertia of neutron stars to the equation of state of neutron-rich matter. {\em Phys. Rev. C} {\bf 2010}, {\em 82}, 025810.
\bibitem[Read(2009)]{Read09} Read, J.S.; Lackey, B.D.; Owen, B.J.; Friedman, J.L. Constraints on a phenomenologically parametrized neutron star equation of state. {\em Phys. Rev. D} {\bf 2009}, {\em 79}, 124032.
\bibitem[Margueron(2018a)]{Margueron17a} Margueron, J.; Hoffmann Casali, R.; Gulminelli, F. Equation of state for dense nucleonic matter from metamodeling. I. Foundational aspects.  {\em Phys. Rev. C} {\bf 2018}, {\em 97}, 025805.
\bibitem[Margueron(2018b)]{Margueron17b} Margueron, J.; Hoffmann Casali, R.; Gulminelli, F. Equation of state for dense nucleonic matter from metamodeling. II. Predictions for neutron star properties. {\em Phys. Rev. C} {\bf 2018}, {\em 97}, 025806.
\bibitem[Annala(2018)]{Annala18} Annala, E.; Gorda, T.; Kurkela, A.; Vuorinen, A. Gravitational-wave constraints on the neutron star matter Equation of State. {\em Phys. Rev. Lett.} {\bf 2018}, {\em 120}, 172703.
\bibitem[Zhang(2018)]{Zhang18} Zhang, N.B.; Li, B.A.; Xu, J. Combined constraints on the equation of state of dense neutron-rich Matter from terrestrial nuclear experiments and observations of Neutron Stars. {\em Astrophys. J.} {\bf 2018}, {\em 859}, 90.
\bibitem[Xie(2019)]{Xie19} Xie, W.J.; Li, B.A. Bayesian inference of high-density nuclear symmetry energy from radii of canonical neutron stars. {\em Astrophys. J.} {\bf 2019}, {\em 883}, 174.
\bibitem[Zhang(2019)]{Zhang19} Zhang, N.B.; Li, B.A. Extracting nuclear symmetry energies at high densities from observations of neutron stars and gravitational waves. {\em Euro. Phys. J. A} {\bf 2019}, {\em 55}, 39.
\bibitem[Zhang(2021)]{Zhang2021} Zhang, N.B.; Li, B.A. Impact of NICER's radius measurement of PSR J0740+6620 on nuclear symmetry energy at suprasaturation densities. {\em Astrophys. J.} {\bf 2021}, {\em 921}, 111.
\bibitem[Oppenheimer(1939)]{Oppenheimer39} Oppenheimer, J.R.; Volkoff, G.M. On massive neutron cores. {\em Phys. Rev.} {\bf 1939}, {\em 55}, 374.
\bibitem[Bombaci(1991)]{Bombaci91} Bombaci, I.; Lombardo, U. Asymmetric nuclear matter equation of state. {\em Phys. Rev. C} {\bf 1991}, {\em 44}, 1892.
\bibitem[Zhang(2019a)]{Zhang19a} Zhang, N.B.; Li, B.A. Delineating effects of nuclear symmetry energy on the radii and tidal polarizabilities of neutron stars. {\em J. Phys. G: Nucl. Part. Phys.} {\bf 2019}, {\em 46}, 014002.
\bibitem[Zhang(2019b)]{Zhang19b} Zhang, N.B.; Li, B.A. Implications of the mass $M=2.17^{+0.11}_{-0.10} M_\odot$ of PSR J0740+6620 on the equation of state of super-dense neutron-rich nuclear matter. {\em Astrophys. J.} {\bf 2019}, {\em 879}, 99.
\bibitem[Zhang(2020)]{Zhang2020} Zhang, N.B.; Li, B.A. Constraints on the Muon Fraction and Density Profile in Neutron Stars. {\em Astrophys. J.} {\bf 2020}, {\em 893}, 61.
\bibitem[Zhang(2020b)]{Zhang2020b} Zhang, N.B.; Qi, B.; Wang, S.Y. Key factor for determining relation between radius and tidal deformability of neutron stars: Slope of symmetry energy. {\em Chin. Phys. C} {\bf 2020}, {\em 44}, 064103.
\bibitem[Zhang(2023)]{Zhang22} Zhang, N.B.; Li, B.A. Impact of symmetry energy on sound speed and spinodal decomposition in dense neutron-rich matter. {\em Euro. Phys. J. A} {\bf 2023}, {\em 59}, 86.
\bibitem[Zhang(2024)]{Zhang24} Zhang, N.B.; Li, B.A. Impact of the nuclear equation of state on the formation of twin stars. \emph{arXiv} \textbf{2024}, arXiv:2406.07396.
\bibitem[Xie(2024)]{Xie24} Xie, W.J.; Li, B.A.; Zhang, N.B. Impact of the newly revised gravitational redshift of X-ray burster GS 1826-24 on the equation of state of supradense neutron-rich matter. {\em Phys. Rev. D} {\bf 2024}, {\em 110}, 043025.
\bibitem[Garg(2018)]{Garg18} Garg, U., Col\`{o}, G. The compression-mode giant resonances and nuclear incompressibility. {\em Prog. Part. Nucl. Phys.} {\bf 2018}, {\em 101}, 55--95.
\bibitem[Shlomo(2006)]{Shlomo06} Shlomo, S.; Kolomietz, V.M.; Col\`{o}, G. Deducing the nuclear matter incompressibility coefficient from data on isoscalar compression modes. {\em Euro. Phys. J. A} {\bf 2006}, {\em 30}, 23--30.
\bibitem[Li(2013)]{Li13} Li, B.A.; Han, X. Constraining the neutron-proton effective mass splitting using empirical constraints on the density dependence of nuclear symmetry energy around normal density. {\em Phys. Lett. B} {\bf 2013}, {\em 727}, 276-281.
\bibitem[Grams(2022)]{Grams22} Grams, G.; Somasundaram, R.; Margueron, J.; Khan, E. Nuclear incompressibility and speed of sound in uniform matter and finite nuclei. {\em Phys. Rev. C} {\bf 2022}, {\em 106}, 044305.
\bibitem[Mondal(2017)]{Mondal17} Mondal, C.; Agrawal, B.K.; De, J.N.; Samaddar, S.K.; Centelles, M.; Vi\~{n}as, X. Interdependence of different symmetry energy elements. {\em Phys. Rev. C} {\bf 2017}, {\em 96}, 021302.
\bibitem[Somasundaram(2021)]{Somasundaram21} Somasundaram, R.; Drischler, C.; Tews, I.; Margueron, J. Constraints on the nuclear symmetry energy from asymmetric matter calculations with chiral NN and 3 N interactions. {\em Phys. Rev. C} {\bf 2021}, {\em 103}, 045803.
\bibitem[Xie(2020)]{Xie20} Xie, W.J.; Li, B.A. Bayesian inference of the symmetry energy of superdense neutron-rich matter from future radius measurements of massive neutron stars. {\em Astrophys. J.} {\bf 2020}, {\em 899}, 4.
\bibitem[Cai(2017)]{Cai17} Cai, B.J.; Chen, L.W. Constraints on the skewness coefficient of symmetric nuclear matter within the nonlinear relativistic mean field model. {\em Nucl. Sci. Tech.} {\bf 2017}, {\em 28}, 185.
\bibitem{Zhang17} Zhang, N.B.; Cai, B.J.; Li, B.A.; Newton, W.G.; Xu, J. How tightly is the nuclear symmetry energy constrained by a unitary Fermi gas?. {\em Nucl. Sci. Tech.} {\bf 2017}, {\em 28}, 181.
\bibitem[Tews(2017)]{Tews17} Tews, I.; Lattimer, J.M.; Ohnishi, A.; Kolomeitsev, E.E. Symmetry parameter constraints from a lower bound on neutron-matter energy. {\em Astrophys. J.} {\bf 2017}, {\em 848}, 105.
\bibitem[Negele(1973)]{Negele73} Negele, J.W.; Vautherin, D. Neutron star matter at sub-nuclear densities. {\em Nucl. Phys. A} {\bf 1973}, {\em 207}, 298--320.
\bibitem[Baym(1971)]{Baym71b} Baym, G.; Pethick, C.; Sutherland, P. The ground state of matter at high densities: Equation of state and stellar models. {\em Astrophys. J.} {\bf 1971}, {\em 170}, 299.
\bibitem[Tolman(1934)]{Tolman34} Tolman, R.C. Effect of inhomogeneity on cosmological models. {\em Proc. Natl. Acad. Sci. USA} {\bf 1934}, {\em 20}, 169--176.
\bibitem[Kubis(2004)]{Kubis04} Kubis, S. Diffusive instability of a kaon condensate in neutron star matter. {\em Phys. Rev. C} {\bf 2004}, {\em 70}, 065804.
\bibitem[Kubis(2007)]{Kubis07} Kubis, S. Nuclear symmetry energy and stability of matter in neutron stars. {\em Phys. Rev. C} {\bf 2007}, {\em 76}, 025801.
\bibitem[Lattimer(2007)]{Lattimer07} Lattimer, J.M.; Prakash, M. Neutron star observations: Prognosis for equation of state constraints. {\em Phys. Rep.} {\bf 2007}, {\em  442}, 109--165.
\bibitem[Xu(2009)]{Xu09} Xu, J.; Chen, L.W.; Li, B.A.; Ma, H.R. Nuclear constraints on properties of neutron star crusts. {\em Astrophys. J.} {\bf 2009}, {\em 697}, 1549.
\bibitem[Ducoin(2011)]{Ducoin11} Ducoin, C.; Margueron, J.; Provid\^{e}ncia, C.; Vidana, I. Core-crust transition in neutron stars: Predictivity of density developments. {\em Phys. Rev. C} {\bf 2011}, {\em 83}, 045810.
\bibitem[Piekarewicz(2014)]{Piekarewicz14} Piekarewicz, J.; Fattoyev, F.J.; Horowitz, C.J. Pulsar glitches: The crust may be enough. {\em Phys. Rev. C} {\bf 2014}, {\em 90}, 015803.
\bibitem[Routray(2016)]{Routray16} Routray, T.R.; Vinas, X.; Basu, D.N.; Pattnaik, S.P.; Centelles, M.; Robledo, L.B.; Behera, B. Exact versus Taylor-expanded energy density in the study of the neutron star crust-core transition. {\em J. Phys. G Nucl. Part. Phys.} {\bf 2016}, {\em 43}, 105101.
\bibitem[Worley(2008)]{Worley} Worley, A.; Krastev, P.G.;Li, B.A. Nuclear constraints on the moments of inertia of neutron stars. {\em Astrophys. J.} {\bf 2008}, {\em 685}, 390.
\bibitem[Xu(2009)]{Xu09b} Xu, J.; Chen, L.W.; Li, B.A.; Ma, H.R. Locating the inner edge of the neutron star crust using terrestrial nuclear laboratory data. {\em Phys. Rev. C} {\bf 2009}, {\em 79}, 035802.
\bibitem[Fortin(2016)]{Fortin} Fortin, M.; Provid\^{e}ncia, C.; Raduta, A.R.; Gulminelli, F.; Zdunik, J.L.; Haensel, P.; Bejger, M. Neutron star radii and crusts: Uncertainties and unified equations of state. {\em Phys. Rev. C} {\bf 2016}, {\em 93}, 035804.
\bibitem[Carreau(2019)]{Carreau19} Carreau, T.; Gulminelli, F.; Margueron, J. Bayesian analysis of the crust-core transition with a compressible liquid-drop model. {\em Euro. Phys. J. A} {\bf 2019}, {\em 55}, 188.
\bibitem[Davis(2024)]{Davis24} Davis, P.J.; Thi, H.D.; Fantina, A.F.; Gulminelli, F.; Oertel, M.; Suleiman, L. Inference of neutron-star properties with unified crust-core equations of state for parameter estimation. {\em Astron. Astrophys.} {\bf 2024}, {\em 687}, A44.
\bibitem[Tsaloukidis(2019)]{Tsaloukidis19} Tsaloukidis, L.; Margaritis, Ch.; Moustakidis, Ch. C. Effects of the equation of state on the core-crust interface of slowly rotating neutron stars. {\em Phys. Rev. C} {\bf 2019}, {\em 99}, 015803.
\end{thebibliography}
\end{document}